\documentclass[authoryear,final,5p,times,twocolumn, colorlinks=true,citecolor=blue,linkcolor=blue]{elsarticle} 
\usepackage{amssymb}
\usepackage{amsmath}
\usepackage{xcolor}
\usepackage{orcidlink} 
\usepackage{soul}

\journal{Journal of High Energy Astrophysics}

\begin{document}

\begin{frontmatter}

\title{Curvature-Radiation Geometries Across the Second CHIME/FRB Fast Radio Burst Population}

\author[label1]{Thonimar V. Alencar\corref{cor1}\orcidlink{0000-0001-8983-8413}}
\ead{thonimar.souza@ufes.br}
\author[label2,label3]{J{\'e}ferson A. S. Fortunato\corref{cor1}\orcidlink{0000-0001-7983-1891}}
\ead{jeferson.fortunato@uct.ac.za}
\author[label1,label4]{Wiliam S. Hipólito-Ricaldi\corref{cor1}\orcidlink{0000-0002-1748-553X}}
\ead{wiliam.ricaldi@ufes.br}
\cortext[cor1]{Corresponding authors}

\affiliation[label1]{organization={Grupo de Física Teórica e Computacional, Departamento de Ciências Naturais, CEUNES, Universidade Federal do Espírito Santo (UFES)},
addressline={Rodovia Governador Mário Covas, Km 60},
city={São Mateus},
postcode={29932-540},state={ES},
country={Brasil}}

\affiliation[label2]{organization={High Energy Physics, Cosmology and 
             Astrophysics Theory (HEPCAT) Group, Department of Mathematics 
             and Applied Mathematics, University of Cape Town},
             addressline={Rondebosch},
             city={Cape Town},
             postcode={7700},
             country={South Africa}}

\affiliation[label3]{organization={African Institute for Mathematical Sciences}, addressline={6 Melrose Road, Muizenberg, Cape Town}, postcode={7945}, country={South Africa}}

\affiliation[label4]{organization={N\'ucleo Cosmo-UFES, CCE, Universidade Federal do Esp\'{\i}rito Santo (UFES)}, addressline={Av. Fernando Ferrari, 540}, city={Vitória},
             postcode={29.075-910},
             country={Brasil}}

\date{Received: \today}

\begin{abstract}
We present a population-level spectral analysis of fast radio bursts from the second CHIME/FRB catalog using three curvature-radiation--motivated templates: point-source, one-dimensional bunch, and paired-bunch cavity, {together with two non-curvature oscillatory baselines}. Fits are evaluated with reduced chi-squared $\chi^2_r$, {the Akaike information criterion (AIC), the Bayesian information criterion (BIC)}, and the Ljung--Box residual autocorrelation test. All three templates yield median $\chi^2_r$ values close to unity for both repeating and non-repeating bursts. Repeaters show narrower $\chi^2_r$ distributions than non-repeaters, with statistically significant but modest population-level differences. {AIC favours the one-dimensional bunch model for the largest fraction of sources
within the curvature-radiation family; BIC also favours the 1d model in both
FRB classes, though with a markedly increased relative share for the simpler
point-source model.} 
However, residual autocorrelation remains widespread across all models: {only $12\%$--$22\%$} of sources simultaneously satisfy goodness-of-fit and residual-independence criteria, indicating persistent structured residuals beyond the tested templates.  These results suggest that while curvature-radiation–motivated geometries capture the dominant spectral envelope of many FRBs, additional physical ingredients or spectral components are required to describe the fine-scale spectral structure of the data. {Comparison with non-curvature baselines shows that oscillatory modulation is the dominant spectral signature within the CHIME band, with the curvature-radiation envelope not independently constrainable at these frequencies.}. The inferred {effective spatial scales} are $\sim$16--18 cm for the one-dimensional model and $\sim$25--28 cm for the cavity model. {Their interpretation as bunch lengths or cavity separations is conditional on the assumed emission model.}
\end{abstract}


\end{frontmatter}

\section{Introduction}

Fast Radio Bursts (FRBs) are millisecond-duration bursts of radio emission  typically originating  from extragalactic sources. Their extreme brightness temperatures, often exceeding \(10^{35}\) K, require coherent radiation processes~\citep[{see the review by}][]{Zhang2023}. Proposed mechanisms include synchrotron maser emission in relativistic shocks, coherent magnetospheric processes such as curvature radiation, and other plasma-based emission scenarios, although the first two are generally considered among the leading classes of models~\citep{Zhang2023}. However, no single mechanism is yet established as dominant, and different models may operate in distinct sub-populations or physical environments. 

{These mechanisms are commonly grouped into two broad classes: coherent magnetospheric emission, including curvature radiation from relativistic charge bunches, in which the radiation originates within or close to the neutron-star magnetosphere; and synchrotron-maser emission at relativistic shocks, in which the radiation is produced well outside the magnetosphere as ejecta from a central engine decelerates in the surrounding medium \citep{Zhang2023}. The low radio efficiency commonly inferred for synchrotron-maser emission can impose demanding energy requirements. For example, a cumulative radio energy of approximately $3\times10^{41}$ erg \citep{Wang2022NatCo}, combined with an efficiency below $10^{-5}$ \citep{Metzger2019}, would imply an engine energy exceeding approximately $3\times10^{46}$ erg. This estimate is subject to assumptions about beaming, bandwidth, the extrapolation of the observed activity level, and the connection between individual bursts and the underlying energy reservoir. The normalized spectral analysis presented here does not constrain the absolute radiative efficiency and therefore cannot discriminate between curvature-radiation and maser scenarios on energetic grounds. Curvature radiation is selected in this work because it provides a particularly tractable analytic connection between an assumed coherent-emission geometry and the resulting spectral shape. This choice does not imply that the present analysis uniquely distinguishes curvature radiation from other intrinsic or propagation-based mechanisms.} 

Among these mechanisms, curvature radiation from relativistic charge bunches within neutron star magnetospheres remains a viable and actively investigated scenario, particularly in compact, highly magnetized environments. {This mechanism has been proposed as one possible explanation for several key observational properties of some FRBs, including their extreme brightness and coherence~\citep{Kumar2017,Lu2018}, strong polarization signatures that may be compatible with highly ordered magnetospheric magnetic fields~\citep{Qu2023}, and GHz-frequency emission consistent with observations~\citep{Liu2023}. It is also compatible with the millisecond variability of the bursts and with the magnetar environments in which at least some sources reside~\citep{Yang2018, Desvignes2024}.} Recent observations~\citep{Mckinven2025,Nimmo2025} further suggest that the emission region is extremely compact, likely located within or just beyond the neutron star magnetosphere, thereby placing important constraints on the characteristic spatial scales of viable emission models. Such compactness is broadly consistent with curvature-radiation scenarios, in which strong magnetic fields and extreme plasma conditions enable coherent emission through charge bunching~\citep{Kumar2017,Lu2018, Wang2022}, although alternative shock-driven interpretations remain viable. However, several theoretical challenges remain, including the formation and survival of coherent charge bunches under realistic magnetospheric conditions and the ability of simple geometrical templates to reproduce the observed spectral diversity of FRBs.

The emission spectrum of curvature radiation from a charged-particle bunch is strongly influenced by the spatial configuration of the emitting charges. Previous studies~\citep{Yang2018,Yang2018b,Yang2020,Yang2023} have shown that different charge-bunch geometries yield distinct spectral behaviors. When the bunch is approximated as a point source, i.e., when its characteristic length is much smaller than the wavelength of the emitted radiation, the underlying theoretical spectrum follows a power-law  with an exponential cutoff~\citep{Yang2018,Yang2018b}. For a one-dimensional, extended bunch, the spectrum is modulated by a sinc--squared function, introducing oscillatory features that depend on the bunch length~\citep{Yang2018}. In configurations involving a bunch-cavity pair, possibly arising from plasma fluctuations, the spectrum exhibits sinusoidal modulations  associated with interference between spatially separated emission regions~\citep{Yang2020}. 

In this work, we present a systematic statistical analysis of FRB spectra using curvature--radiation--motivated phenomenological templates, applied to 4536 FRBs from the second CHIME/FRB public data release~\citep{abbott_second_2026}. To our knowledge, this is the largest homogeneous FRB sample analysed within this specific template framework to date. Each spectrum is extracted directly from the HDF5 files and fitted via nonlinear least squares under physically motivated constraints. While standard goodness-of-fit metrics quantify global agreement, they do not capture structured residual deviations. To address this limitation, we additionally employ the Ljung–Box test on residuals \citep{BoxPierce1970, LjungBox1978, Mokeddem2023}, providing a complementary diagnostic of residual structure beyond standard goodness-of-fit metrics. Our goal is not to establish curvature radiation as a unique emission mechanism, but rather to assess, at the population level, how well simple geometrical models of coherent curvature radiation reproduce the observed FRB spectra and where they systematically fail.

The paper is organized as follows.  Section~\ref{sec:radiation}  introduces the theoretical models,  Section~\ref{sec:application}  describes the data and methodology, Section~\ref{sec:results} presents the results and their physical interpretation, and Section~\ref{sec:conclusions} summarizes the main conclusions.

\section{{Curvature Radiation from Relativistic Charge Bunches}}\label{sec:radiation}

Coherent curvature radiation from relativistic charge bunches is considered a plausible mechanism for producing  the observed spectral structures in FRBs~\citep{Zhang2023}. The emitted power depends sensitively on the geometry and spatial coherence of the emitting ensemble, which modulate the intrinsic single-particle curvature-radiation spectrum through collective interference effects~\citep{Yang2023}. 

For a relativistic particle ensemble containing $\mathcal{N}$ charges, the total spectral intensity can be expressed as~\citep{Yang2018}
\begin{equation}
     S(\nu)  \approx F_\nu(\mathcal{N})\,\frac{d^2 I}{d\nu\,d\Omega}
\end{equation}
where $F_{\nu}(\mathcal{N})$ is the coherence enhancement factor, determined by the spatial distribution of the charge bunch, and $d^2 I/d\nu d\Omega$ is the single-particle curvature spectrum, as described in \ref{appendix:single_particle}. In the fully coherent limit, $F_\nu(\mathcal{N}) = \mathcal{N}^2$, while partial coherence introduces a frequency-dependent suppression governed by the characteristic spatial scales of the system.

To connect these coherence properties with observable spectral signatures, we consider three representative geometrical configurations of the emitting bunches, described by idealized  spectral models derived from the curvature-radiation formalism.

\paragraph{(i) Point-like bunch}
When the spatial extent of the bunch is much smaller than the wavelength, all charges radiate coherently with $F_{\nu}(\mathcal{N}) = \mathcal{N}^2$. In this limit, the spectrum reduces to
\begin{equation}\label{eq:point}
    S_{\text{point}}(\nu) \approx A\,\gamma^2 \mathcal{N}^2 \left( \frac{\nu}{\nu_c} \right)^{2/3} e^{-\nu/\nu_c},
\end{equation}
where $A$ is a normalization constant defined in \ref{appendix:single_particle}, absorbing dimensional prefactors, $\gamma$ is the Lorentz factor, and $\nu_c = 3c\gamma^3/(4\pi\rho)$ denotes the critical frequency associated with a magnetic field line of curvature radius $\rho$~\citep{Jackson1999}.

\paragraph{(ii) One-dimensional bunch}
For a bunch of longitudinal length $l$, phase coherence is progressively reduced at frequencies above  $\nu_l = c/(\pi l)$, while introducing an oscillatory interference pattern:
\begin{equation}\label{eq:1d}
    S_{\text{1D}}(\nu) \approx A\,\gamma^2 \mathcal{N}^2 \left( \frac{\nu}{\nu_c} \right)^{2/3} 
    \text{sinc}^2\!\left( \frac{\nu}{\nu_l} \right) e^{-\nu/\nu_c}.
\end{equation}
This behavior reflects the gradual loss of phase coherence along the longitudinal direction of the bunch.
\paragraph{(iii) Bunch-cavity pair}
If the emission arises from a two-source configuration, such as a bunch–cavity pair separated by a distance $d$, the interference pattern  becomes sinusoidal, with modulation frequency $\nu_d = c/(\pi d)$. In this case, the spectrum takes the form
\begin{equation}\label{eq:cavity}
    S_{\text{cavity}}(\nu) \approx A\,\gamma^2 \mathcal{N}^2 \left( \frac{\nu}{\nu_c} \right)^{2/3} 
    \sin^2\!\left( \frac{\nu}{\nu_d} \right) e^{-\nu/\nu_c}.
\end{equation}
Such configurations may phenomenologically represent interference between spatially separated coherent emitting regions, potentially associated with plasma inhomogeneities or pair-production regions within the magnetosphere. 

In the point-like spectrum, the peak frequency satisfies $\nu_p=(2/3)\nu_c$, with $\nu_c \propto \gamma^3/\rho$ providing a connection between the observed spectral peak and the underlying physical parameters. In the presence of interference terms (Eqs.~\ref{eq:1d} and~\ref{eq:cavity}),  multiple local maxima may appear, although the overall spectral envelope remains controlled by $\nu_c$. However, as discussed below, $\nu_c$ is not independently constrained within the CHIME frequency range, and therefore $\gamma$ and $\rho$ cannot be directly inferred in the phenomenological analysis performed here.

These three configurations therefore define distinct and, in principle, testable spectral signatures that can be directly confronted with observational data.

\section{Data Processing, Spectral Modeling, and Statistical Framework}
\label{sec:application}

\subsection{Data Preprocessing}\label{sec:preprocessing}

The curvature-radiation templates were fitted to individual FRB spectra from the second CHIME/FRB public data release~\citep{abbott_second_2026}. Each event is provided as an HDF5 (\texttt{.h5}) file, in which the processed Stokes--I dynamic spectrum is stored in the dataset \texttt{data}, together with associated metadata. As described in the metadata, the dynamic spectrum $ I(\nu,t)$ is mean-subtracted and mean-normalized on a per-frequency-channel basis using the channel-wise mean computed over the full time series, incoherently dedispersed to a fixed reference frequency, and restricted to a temporal window centered on the mean arrival time of the identified sub-bursts~\citep{abbott_second_2026}. From this dynamic spectrum, the resulting dimensionless spectral profile is obtained as $S_i(\nu)=\langle I_i(\nu,t)\rangle_t$, where the index $i$ labels individual bursts in the CHIME/FRB Catalog~2, comprising 4536 events used in this analysis.\footnote{Although Catalog~2 contains 4539 events, FRB20190422B, FRB20190517D, and FRB20190415C are not included in the public data release and were therefore excluded.}

To ensure homogeneous and reproducible preprocessing across the sample, each spectrum was filtered to remove frequency channels affected by radio-frequency interference (RFI) or invalid values. RFI excision was performed in two stages. First, frequency channels flagged by the CHIME/FRB pipeline as missing or RFI-contaminated were removed using the per-event \texttt{good\_freq} boolean mask provided in the public data release \citep{abbott_second_2026}, thereby excluding channels identified a priori as problematic by the instrument pipeline. Second, a supplementary robust masking step was applied to the remaining channels using an interquartile range (IQR) criterion, excluding channels with spectral amplitude outside $[Q_1 - f_{\mathrm{IQR}}\,\mathrm{IQR},\, Q_3 + f_{\mathrm{IQR}}\,\mathrm{IQR}]$, supplemented by a median absolute
deviation (MAD) criterion~\citep{morello_iqrm_2021,buch_realtime_2019}, excluding channels satisfying $|S_i - \tilde{S}| > f_{\mathrm{MAD}}\cdot\mathrm{MAD}$, where $\tilde{S}$ is the median spectral amplitude. Both thresholds were fixed uniformly across the full sample ($f_{\mathrm{IQR}} = 1.5$ and $f_{\mathrm{MAD}} = 5.0$) to preserve methodological consistency across all bursts. Across the 4536-event sample, the combined masking procedure removed a median fraction of $\sim0.8\%$ of channels per burst (interquartile range $0.7$--$1\%$), indicating that the masking remains conservative while providing an additional safeguard against residual contamination. 

After masking, frequency rebinning was applied by averaging over contiguous groups of channels, controlled by a binning factor $f_{\mathrm{bin}} = 2^n$ for $n = 0, 1, \ldots, 8$. This multiscale rebinning serves both to reduce channel-level noise and to assess the robustness of the fitted spectral behavior across different effective frequency resolutions. Since the unbinned spectra contain exactly $16\,384$ (or $2^{14}$) channels~\citep{Amiri2018,Amiri2021,Amiri2022}, all values divide the dataset exactly, ensuring uniform weighting. 

The rebinned spectra were computed using a NaN-aware mean to maintain robustness against any remaining flagged or invalid samples. Partially masked bins contribute only through their valid channels. The resulting rebinned frequency array $\nu_i$ and spectral amplitude $S_i$ were used as inputs to the fitting procedure. This preprocessing pipeline minimizes contamination from instrumental artifacts while preserving spectral features relevant for model testing.

\subsection{Spectral Model Fitting and Goodness-of-Fit Diagnostics} \label{sec:fitting}
\medskip
Model fitting was performed using a nonlinear least-squares minimization applied to the three curvature-radiation–motivated spectral templates: a \texttt{point} bunch model [Eq.~\eqref{eq:point}], a one-dimensional (\texttt{1D}) extended bunch model [Eq.~\eqref{eq:1d}], and a \texttt{bunch-cavity} pair model [Eq.~\eqref{eq:cavity}]. The point-like model contains one free parameter, $s_0$, corresponding to an overall amplitude normalization, whereas the one-dimensional and cavity models contain two free parameters, $(s_0,b)$. In these latter models, $b$ sets the characteristic interference scale controlling the spectral modulation, with $b=1/\nu_\ell$ for the one-dimensional model and $b=1/\nu_d$ for the cavity model. Because the point-like model contains no interference term, $b$ is not defined in this case. Consequently, both AIC and BIC penalize the two-parameter models more strongly. The fitted parameters should therefore be interpreted as effective template parameters rather than as a complete microscopic description of the emitting plasma. {Throughout this work $b$ is referred to as an effective modulation timescale, and its interpretation in terms of a specific emitting geometry is treated as conditional on the assumed template (Section \ref{sec:physical_constraints}).}

The exponential cutoff factor $e^{-\nu/\nu_c}$ present in the full curvature-radiation spectrum was omitted from the fitted templates because $\nu_c$ is not independently identifiable within the CHIME/FRB observing band (400–800 MHz), being nearly degenerate with the amplitude normalization $s_0$. Exploratory fits including this term typically converged to parameter boundaries or produced unstable covariance matrices, indicating numerical non-identifiability rather than physical absence of a cutoff.  Omitting $e^{-\nu/\nu_c}$ is not expected to strongly bias the inferred interference scale $b$ , because the cutoff varies slowly compared with the fitted modulation scale across the CHIME band. A quantitative assessment of this approximation is deferred to future broadband analyses where $\nu_c$ becomes independently identifiable.

Each model was fitted to the observed spectra via non-linear least-squares minimization using the \texttt{least\_squares} function from the \texttt{scipy.optimize} library \citep{Virtanen2020}. Bound-constrained optimization was performed with the trust-region reflective algorithm, and the Jacobian was approximated using a two-point finite-difference scheme. The amplitude parameter $s_0$ was restricted to the interval $[0,1]$, consistent with the normalized amplitude distribution of the catalog spectra. The interference scale $b$ was bounded within $[10^{-4}, 10^{-1}]$~$\mu$s. Using  $d = c / (\pi \nu_d) = c\, b / \pi$ and $L = c / (\pi \nu_l) = c\, b / \pi$ for the cavity separation $d$ and longitudinal extent $L$, respectively, this interval corresponds to characteristic spatial scales $L \sim c b / \pi$  ranging from $0.95$~cm to $9.55$~m. These values are broadly consistent with theoretical expectations for microscopic coherent radio emission at $\sim1$~GHz~\citep{Kumar2017,Cui2023}, where typical bunch sizes and charge-separation scales are of the order $\sim 10-12$~cm~\citep{Cui2023,Yang2020}. The recovered median {effective spatial scales} reported in Section \ref{sec:physical_constraints} lie comfortably within these bounds ($[10^{-4},10^{-1}]$~$\mu$s){, although for the \texttt{cavity} model the lower quartile of $b$ coincides with the lower bound of this interval (Table~\ref{tab:physical_constraints})}.

Weights were estimated from the per-channel noise provided in the CHIME/FRB data release. For each frequency channel $i$, the raw standard deviation $\sigma_i^{\mathrm{raw}} = \sqrt{\langle(I_i - \langle I_i\rangle)^2\rangle_t}$ was normalised by the channel mean to yield the dimensionless relative noise $\hat{\sigma}_i = \sigma_i^{\mathrm{raw}} / \langle I_i \rangle$, with channels satisfying $|\langle I_i\rangle| \leq 10^{-10}$ excluded. This was further propagated through time-averaging as $\sigma_i = \hat{\sigma}_i / \sqrt{N_i}$, where $N_i = \sum_t f_{i,t}$ is the number of unflagged samples ($f_{i,t}\in\{0,1\}$); channels with $N_i = 0$ were also excluded. The heteroscedastic weight $w_i = \sigma_i^{-1}$ captures frequency-dependent noise variations across the CHIME band without contamination from the burst signal itself. The residual vector minimised in the fit is $r_i = w_i\,[S_i - S_{\mathrm{model}}(\nu_i;\,s_0,\,b)]$, and $\chi^2$ and $\chi^2_r$ were computed from the weighted residuals accordingly:
\begin{align}
\chi^2 &= \sum_i r_i^2, \\[3pt]
\chi^2_r &= \frac{\sum_i r_i^2}{N_{\mathrm{data}} - N_{\mathrm{par}}}
\end{align}
where $N_{\mathrm{data}}$ is the number of spectral channels surviving the full masking chain (pipeline mask, IQR/MAD excision, and $\sigma_i$ exclusions) and $N_{\mathrm{par}}$ is the number of free model parameters. 

Assuming independent Gaussian noise for likelihood construction, while residual autocorrelation is evaluated separately as a diagnostic of this assumption, the log-likelihood is
\begin{equation}
\ln\mathcal{L} = -\frac{1}{2}\sum_i \left[r_i^2 + \ln(2\pi\sigma_i^2)\right],
\end{equation}
and the information criteria are computed as
\begin{align}
\mathrm{AIC} &= -2\ln\mathcal{L} + 2N_{\mathrm{par}}, \\[3pt]
\mathrm{BIC} &= -2\ln\mathcal{L} + N_{\mathrm{par}}\ln N_{\mathrm{data}}.
\end{align}

Since $\sigma_i$ is fixed prior to fitting, the $\ln(2\pi\sigma_i^2)$ term cancels in $\Delta \mathrm{AIC}$ and $\Delta \mathrm{BIC}$ for model comparisons performed on the same burst. 

\begin{figure*}[ht!]
\centering
\includegraphics[width=\textwidth]{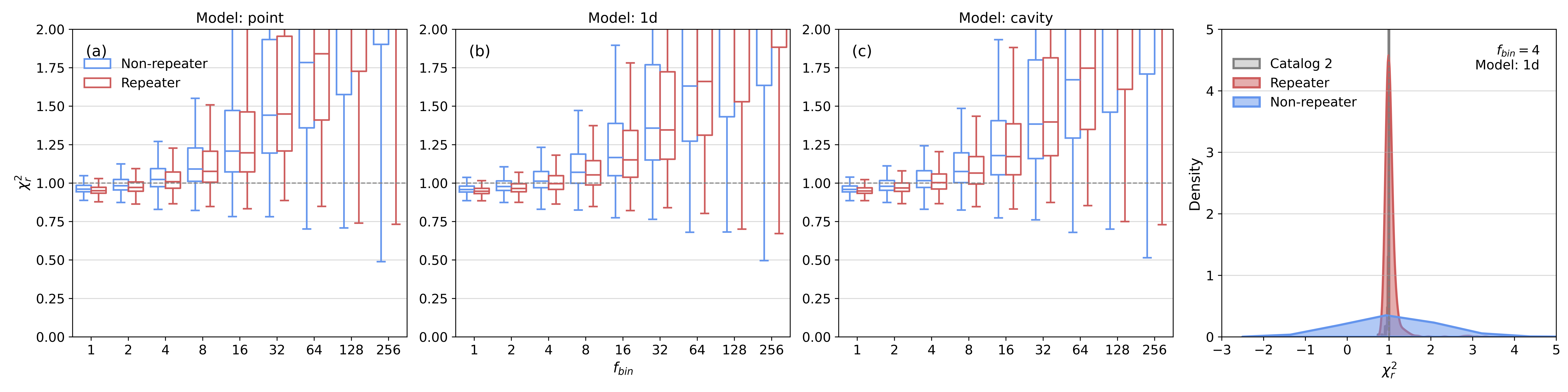}
\caption{Distribution of reduced chi-squared ($\chi^2_r$) values for the three curvature-radiation models. Panels~(a)--(c) show boxplots of $\chi^2_r$ as a function of the frequency-channel binning factor $f_{\rm bin}$ for the \texttt{point}, \texttt{1d}, and \texttt{cavity} models, respectively. Each box shows the median, interquartile range, and overall spread, with outliers 
excluded for clarity, for repeaters (red) and non-repeaters (blue). Panel~(d) shows the KDE of $\chi^2_r$ at $f_{\rm bin} = 4$, including the CHIME catalog reference (gray).}
\label{fig:chi2_distribution}
\end{figure*}
\begin{table*}[ht]
    \centering
    \caption{Pass rate ($\chi^2_r \in [0.75, 1.25]$) and summary statistics 
    at $f_{\rm bin} = 4$.}
    \label{tab:chi2_summary}
    \begin{tabular}{llccccc}
        \hline
        Model & Class & $N$ & $\bar{\chi}^2_r$ & $\tilde{\chi}^2_r$ 
              & $\sigma$ & Pass (\%) \\
        \hline
        \texttt{1d}     & Non-rep. & 3554 & 1.273 & 1.012 & 4.751 & 92.7 \\
                        & Rep.     &  981 & 1.033 & 0.997 & 0.189 & 96.0 \\
        \texttt{cavity} & Non-rep. & 3555 & 1.291 & 1.016 & 5.208 & 92.1 \\
                        & Rep.     &  981 & 1.045 & 1.003 & 0.222 & 95.5 \\
        \texttt{point}  & Non-rep. & 3555 & 1.330 & 1.023 & 5.437 & 90.2 \\
                        & Rep.     &  981 & 1.075 & 1.009 & 0.433 & 94.3 \\
        \hline
    \end{tabular}
\end{table*}

\begin{table*}[ht]
    \centering
    \caption{Summary statistics for AIC and BIC.
             $\bar{x}$ and $\tilde{x}$ denote the mean and median,
             respectively, while $x_{25}$ and $x_{75}$ represent the 25th
             and 75th percentiles.}
    \label{tab:aic_bic_summary}
    \begin{tabular}{llllrrrr}
        \hline
        Model & Criterion & Class & $N$ & $\bar{x}$ & $\tilde{x}$
              & $x_{25}$ & $x_{75}$ \\
        \hline
        \texttt{1d}
            & AIC & Non-rep. & 2\,549 & $-18\,634$ & $-18\,756$
                             & $-20\,181$ & $-17\,250$ \\
            &     & Rep.     &   768 & $-18\,505$ & $-18\,642$
                             & $-19\,683$ & $-17\,181$ \\
            & BIC & Non-rep. & 2\,549 & $-18\,623$ & $-18\,745$
                             & $-20\,169$ & $-17\,239$ \\
            &     & Rep.     &   768 & $-18\,493$ & $-18\,631$
                             & $-19\,671$ & $-17\,170$ \\
        \texttt{cavity}
            & AIC & Non-rep. & 2\,549 & $-18\,627$ & $-18\,750$
                             & $-20\,172$ & $-17\,247$ \\
            &     & Rep.     &   768 & $-18\,487$ & $-18\,610$
                             & $-19\,674$ & $-17\,170$ \\
            & BIC & Non-rep. & 2\,549 & $-18\,616$ & $-18\,738$
                             & $-20\,160$ & $-17\,236$ \\
            &     & Rep.     &   768 & $-18\,475$ & $-18\,599$
                             & $-19\,662$ & $-17\,158$ \\
        \texttt{point}
            & AIC & Non-rep. & 2\,549 & $-18\,607$ & $-18\,732$
                             & $-20\,152$ & $-17\,233$ \\
            &     & Rep.     &   768 & $-18\,466$ & $-18\,592$
                             & $-19\,676$ & $-17\,165$ \\
            & BIC & Non-rep. & 2\,549 & $-18\,602$ & $-18\,726$
                             & $-20\,146$ & $-17\,227$ \\
            &     & Rep.     &   768 & $-18\,460$ & $-18\,587$
                             & $-19\,670$ & $-17\,160$ \\
        {\texttt{powerlaw}}
            & AIC & Non-rep. & 2\,614 & $-18\,644$ & $-18\,758$
                             & $-20\,199$ & $-17\,200$ \\
            &     & Rep.     &   703 & $-18\,523$ & $-18\,661$
                             & $-19\,692$ & $-17\,100$ \\
            & BIC & Non-rep. & 2\,614 & $-18\,627$ & $-18\,741$
                             & $-20\,182$ & $-17\,183$ \\
            &     & Rep.     &   703 & $-18\,506$ & $-18\,644$
                             & $-19\,674$ & $-17\,083$ \\
        {\texttt{sinc2}}
            & AIC & Non-rep. & 2\,614 & $-18\,643$ & $-18\,760$
                             & $-20\,201$ & $-17\,202$ \\
            &     & Rep.     &   703 & $-18\,520$ & $-18\,658$
                             & $-19\,685$ & $-17\,095$ \\
            & BIC & Non-rep. & 2\,614 & $-18\,631$ & $-18\,748$
                             & $-20\,190$ & $-17\,191$ \\
            &     & Rep.     &   703 & $-18\,508$ & $-18\,647$
                             & $-19\,673$ & $-17\,084$ \\
        \hline
    \end{tabular}
\end{table*}

\subsection{Residual Autocorrelation Analysis: Ljung--Box Test}
\label{sec:ljungbox}

Good global fits do not necessarily imply that the residuals are statistically independent. To test for residual autocorrelation left by the fitted templates, we applied the Ljung--Box portmanteau test to the weighted post-fit residuals $r_i$ for each burst and model. The null hypothesis assumes no autocorrelation up to a maximum lag $h$.

The Ljung--Box statistic is
\begin{equation}
Q(h) = N_{\mathrm{data}}(N_{\mathrm{data}}+2)\sum_{k=1}^{h}
\frac{\hat{\rho}_k^2}{N_{\mathrm{data}}-k},
\end{equation}
where $N_{\mathrm{data}}$ is the number of unmasked spectral channels and $\hat{\rho}_k$ is the sample autocorrelation at lag $k$ \citep{LjungBox1978,Box2015}. Under the null hypothesis, $Q(h)$ approximately follows a $\chi^2$ distribution with $h$ degrees of freedom.

We implemented the test using the \texttt{acorr\_ljungbox} routine from the \texttt{statsmodels} package \citep{Seabold2010}. We adopt $h=20$, corresponding to short-range correlations over approximately $0.5\%$ of the typical spectral length at the reference binning factor   $f_{\mathrm{bin}}=4$. A fit is considered to pass the residual-independence criterion when $p_{\rm LB}>0.05$.

Because RFI masking and finite-sample effects can affect the strict white-noise interpretation of the test, we use the Ljung--Box $p$-values as approximate diagnostics of residual structure rather than as exact tests of independent Gaussian noise.  Additional tests indicate that masking patterns are frequently non-random, consistent with clustered RFI contamination, although the median masked fraction remains only $\sim0.8\%$ per burst (\ref{app:tables}). Relative pass rates are therefore interpreted as comparative measures of residual structure.

\section{Results}\label{sec:results} 
\subsection{Frequency Rebinning}

Figure~\ref{fig:chi2_distribution} shows the $\chi^2_r$ distributions across all three models. Panels (a)--(c) display boxplots of $\chi^2_r$ as a function of $f_{\rm bin}$ for the \texttt{point}, \texttt{1d}, and \texttt{cavity} models, respectively, separated by FRB class. Panel~(d) shows the kernel density estimate (KDE) of $\chi^2_r$ for repeaters and non-repeaters at $f_{\rm bin} = 4$, together with the CHIME catalog reference distribution. In the first three panels, at $f_{\mathrm{bin}} \leq 4$, reduced chi-squared values remain close to unity with minimal dispersion, while larger binning factors cause the median to progressively deviate from unity with increasing dispersion, reflecting the gradual loss of spectral information associated with excessive smoothing. Across all binning factors, repeater spectra exhibit systematically narrower $\chi^2_r$  distributions than non-repeaters, indicating more homogeneous fit-quality distributions independently of the adopted spectral resolution. 

At $f_{\mathrm{bin}} = 4$, the distributions remain tightly clustered around unity for both repeaters and non-repeaters across all three spectral models [Fig.~\ref{fig:chi2_distribution}(d)]. Furthermore,  Kolmogorov--Smirnov~\citep{Kolmogorov1933, Smirnov1948} and Mann--Whitney~\citep{Wilcoxon1945, MannWhitney1947} tests confirm a statistically significant separation between the two populations ($p_{KS} \sim 10^{-7}-10^{-4}$, $p_{MW} \sim 10^{-10}-10^{-6}$), confirming that the distributional differences between the two populations are statistically significant across all three spectral models (see discussion below and Table~\ref{tab:stat_tests}). 

We therefore adopt $f_{\mathrm{bin}} = 4$ as a reference  binning scale for the remainder of the analysis, as it provides a practical compromise between noise suppression and preservation of spectral structure within the native CHIME frequency resolution.

\subsection{Reduced Chi-Squared Distribution}

Focusing on the reference binning $f_{\rm bin} = 4$, all  three models yield median values close to unity ($\tilde{\chi}^2_r \approx 0.997$--$1.009$ for repeaters; $1.012$--$1.023$ 
for non-repeaters), consistent with broadly acceptable global fits (Table~\ref{tab:chi2_summary}). However, the two populations exhibit noticeably different distribution shapes. Non-repeaters show broader distributions with enhanced variance and extended high $\chi^2_r$ tails ($\bar{\chi}^2_r \approx 1.273$--$1.330$, $\sigma \lesssim 5.437$) compared to repeaters ($\bar{\chi}^2_r \approx 1.033$--$1.075$, $\sigma \lesssim 0.433$), indicating that the broadening is driven by a minority of less accurately fitted events. This contrast indicates a systematic difference in fit quality between the two populations. Pass rates ($\chi^2_r \in [0.75, 1.25]$) are also systematically higher for repeaters ($94.3$--$96.0\%$) than for non-repeaters ($90.2$--$92.7\%$) across all models (Table~\ref{tab:chi2_summary}), with the \texttt{1d} configuration yielding the highest values for both populations. 

The statistical behaviour of the fits is further illustrated by the visual inspection of individual events. Figure~\ref{fig:spectral_fits} (\ref{app:tables}) displays the spectral fit diagnostics for four representative FRBs under the  best-performing \texttt{1d} bunch model. In these examples, the best-fit models (dashed red lines) closely track the binned data, effectively capturing the interference-driven oscillatory features. Residual histograms and Q--Q plots indicate approximate Gaussian behaviour in the central distributions, with mild tail deviations possibly associated with residual structure. This behaviour is broadly consistent with the Gaussian noise assumption adopted in the fitting procedure and remains compatible with the residual autocorrelation effects discussed in  Section~\ref{sec:ljungbox}.

Two-sample Kolmogorov–Smirnov and Mann–Whitney U tests yield highly significant p-values ($p_{KS} \sim 10^{-7}-10^{-4}$, $p_{MW} \sim 10^{-10}-10^{-6}$) across all models, confirming that the $\chi^2_r$ distributions of the two populations are statistically distinguishable (Table~\ref{tab:stat_tests}). However, given the large sample sizes ($N \approx 981-3555$), even small distributional differences produce significant test statistics; the associated effect sizes remain modest ($D \approx 0.080 - 0.103$ for the KS statistic), indicating substantial overlap between the populations. These results therefore indicate a systematic but quantitatively modest difference between repeaters and non-repeaters, rather than a sharp physical dichotomy.  The largest separation is obtained for the \texttt{1d} model ($D = 0.103$), consistent with its superior performance in pass rates and dispersion.

Taken together, these results suggest that repeater spectra are more consistently described by low-dimensional phenomenological templates, possibly reflecting greater regularity in their spectral structure.

\subsection{Model Comparison via AIC and BIC}\label{sec:aicbic}

\begin{figure*}[ht]
    \centering
    \includegraphics[width=\textwidth]{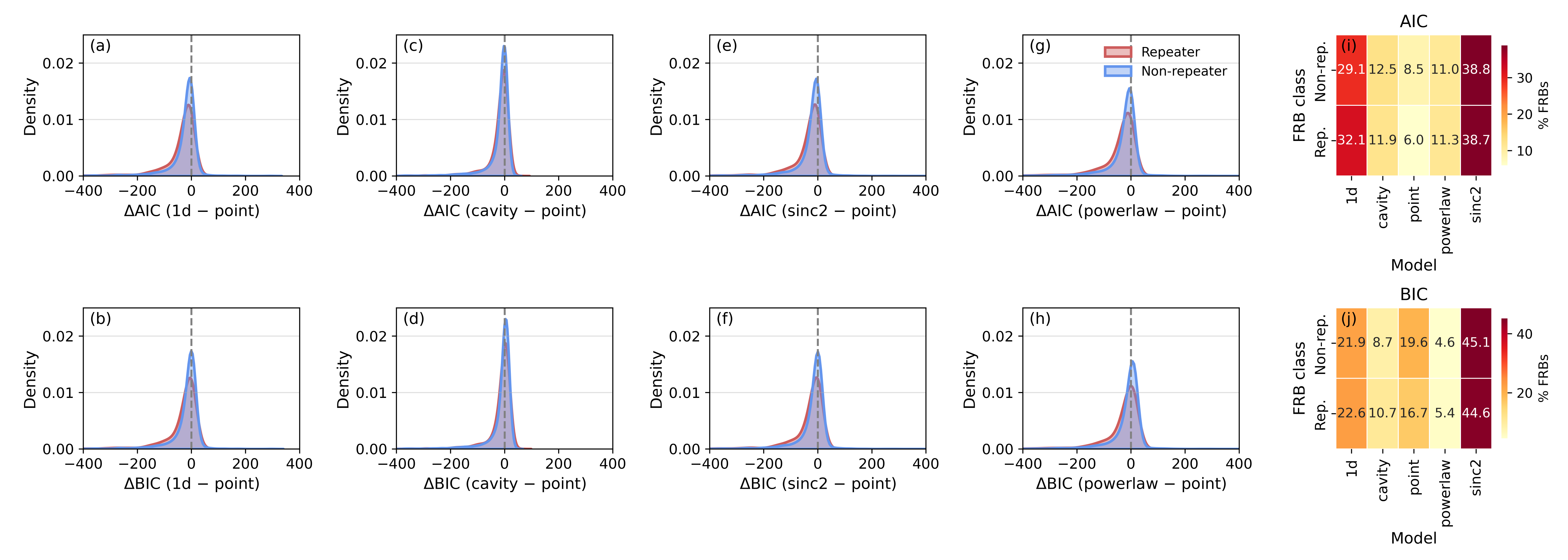}
    \caption{Panels~(a)--(h): kernel density estimates of $\Delta\mathrm{AIC}$
    (top row) and $\Delta\mathrm{BIC}$ (bottom row) relative to the
    \texttt{point} reference model, for each comparison model (columns).
    Negative values indicate that the comparison model is preferred over
    the reference. The vertical dashed line marks $\Delta = 0$;
    distributions to its left favour the comparison model. Colours
    distinguish FRB classes (repeaters and non-repeaters). Panels~(i)--(j):
    fraction of FRBs best fitted by each model according to AIC~(i) and
    BIC~(j), split by FRB class. {In addition to the
    three curvature-radiation-inspired templates, \texttt{point},
    \texttt{1d}, and \texttt{cavity}, the comparison includes two
    non-curvature baseline models, \texttt{sinc2} and \texttt{powerlaw}.}}
    \label{fig:aic_bic_distributions}
\end{figure*}

To enable a consistent model comparison, we define the information-criterion differences relative to the \texttt{point} model, $\Delta\mathrm{AIC} = \mathrm{AIC}_{m} - \mathrm{AIC}_{\mathrm{point}}$ and $\Delta\mathrm{BIC} = \mathrm{BIC}_{m} - \mathrm{BIC}_{\mathrm{point}}$, where $m$ denotes a given comparison model, allowing a direct comparison
of relative model preference within each FRB class. {In addition to the three curvature-radiation-inspired templates considered previously, we include two non-curvature baseline
models. The first is a pure oscillatory profile,
\begin{equation}
S_{\mathrm{sinc2}}(\nu) \propto
\mathrm{sinc}^{2}\left(\frac{\nu}{\nu_l}\right),
\end{equation}
while the second combines the oscillatory component with a power-law spectral envelope,
\begin{equation}
S_{\mathrm{powerlaw}}(\nu) \propto
\nu^{\alpha}
\mathrm{sinc}^{2}\left(\frac{\nu}{\nu_l}\right).
\end{equation}
These two models provide non-curvature baselines that retain the oscillatory spectral structure without imposing the curvature-radiation spectral form, allowing us to test whether the apparent preference for any particular model is specifically associated with the curvature-radiation spectrum or primarily with its oscillatory modulation.}

Throughout this analysis, a quality cut is applied: only fits satisfying
$0.75 \leq \chi^2_r \leq 1.25$ are retained, ensuring that the information
criteria are evaluated exclusively over well-constrained spectral fits.
{For the five-model comparison, we further restrict the
model-selection analysis to FRBs for which all five models provide valid
fits and satisfy the quality criterion. At $f_{\rm bin}=4$, this yields a
common sample of $3317$ FRBs. The fractions shown in Fig.~\ref{fig:aic_bic_distributions}(i)--(j) are
calculated exclusively from this common sample, so that every FRB is
compared using the same set of five models.}

Table~\ref{tab:aic_bic_summary} lists the summary statistics for the absolute
AIC and BIC values. Under the quality cut, the three models yield
remarkably similar median values across both FRB classes. For
non-repeaters, the AIC medians are $-18{,}756$ (\texttt{1d}),
$-18{,}750$ (\texttt{cavity}), and $-18{,}732$ (\texttt{point}), while
for repeaters they are $-18{,}642$, $-18{,}610$, and $-18{,}592$,
respectively.
{With the inclusion of the two non-curvature baselines,
the five models likewise exhibit closely spaced median information
criteria. Under the quality cut, the AIC medians for
non-repeaters are $-18{,}756$ (\texttt{1d}), $-18{,}750$
(\texttt{cavity}), $-18{,}732$ (\texttt{point}), $-18{,}758$
(\texttt{powerlaw}), and $-18{,}760$ (\texttt{sinc2}), while for
repeaters they are $-18{,}642$, $-18{,}610$, $-18{,}592$, $-18{,}661$,
and $-18{,}658$, respectively.}
The BIC values follow an essentially identical pattern, differing from
AIC by at most $\sim!10$ units in all cases.
{The additional baseline models therefore do not introduce
a large separation in the absolute information-criterion values among
the competing templates.}
Minor differences in sample size across models arise from the
model-dependent application of the $\chi^2_r$ quality cut.

The $\Delta$AIC and $\Delta$BIC distributions
[Fig.~\ref{fig:aic_bic_distributions}(a)--(d)] are sharply peaked near
zero, indicating that the three templates provide comparable fits within
the quality-cut sample. Negative tails indicate that the 1d and cavity
templates are preferred over the point model for a subset of sources,
but the overall distributions remain highly overlapping between repeaters
and non-repeaters.
{With the inclusion of the two non-curvature baselines,
the $\Delta\mathrm{AIC}$ and $\Delta\mathrm{BIC}$ distributions remain
concentrated near zero, indicating that several of the five templates
provide comparably good fits within the quality-selected sample. In
particular, the \texttt{sinc2} model can reproduce the oscillatory
spectral structure without the curvature-radiation spectral envelope.
Thus, an improvement relative to the \texttt{point} model cannot by
itself be interpreted as evidence for the curvature-radiation spectral
form. {Moreover, the oscillatory modulation component --- common 
to both \texttt{1d} and \texttt{sinc}$^2$ --- accounts 
for the dominant spectral structure, while the 
curvature-radiation envelope remains degenerate with the 
\texttt{sinc}$^2$ baseline within the CHIME band.}}

Figure~\ref{fig:aic_bic_distributions}(i)--(j) shows the fraction of FRBs
for which each model yields the minimum AIC or BIC.
{For the extended five-model comparison, restricted to the
common sample, the AIC results show that the \texttt{sinc2} model is the
most frequently preferred template for both non-repeaters and repeaters,
accounting for $38.8\%$ and $38.7\%$ of the respective samples. The
\texttt{1d} model is preferred for $29.1\%$ of non-repeaters and $32.1\%$
of repeaters, while \texttt{cavity}, \texttt{point}, and \texttt{powerlaw}
account for the remaining cases ($12.5\%$, $8.5\%$, and $11.0\%$ for
non-repeaters; $11.9\%$, $6.0\%$, and $11.3\%$ for repeaters). Thus, once
the non-curvature baseline models are included, the AIC results no longer
show a dominant preference for the \texttt{1d} curvature-radiation
template.}

{The BIC results, likewise restricted to the common
sample, give a different ranking. For non-repeaters, the \texttt{sinc2}
model is the most frequently selected template ($45.1\%$), followed by
 \texttt{1d} ($21.9\%$), \texttt{point} ($19.6\%$), \texttt{cavity}
($8.7\%$), and \texttt{powerlaw} ($4.6\%$). For repeaters, the same
ordering holds: \texttt{sinc2} ($44.6\%$), \texttt{1d} ($22.6\%$),
\texttt{point} ($16.7\%$), \texttt{cavity} ($10.7\%$), and
\texttt{powerlaw} ($5.4\%$). The different rankings obtained with AIC
and BIC reflect their different penalties for model complexity.
Notably, under BIC the \texttt{1d} template is not the most frequently
selected model for either FRB class -- the non-curvature \texttt{sinc2}
baseline is preferred by a wide margin in both cases.} {The fitted power-law index $\alpha$ is poorly constrained,
with median values near zero and broad interquartile ranges
for both FRB classes, indicating that the \texttt{powerlaw}
model effectively reduces to the \texttt{sinc}$^2$ baseline
in most cases and confirming that the spectral envelope is
not independently constrainable within the CHIME band.}

Kolmogorov--Smirnov and Mann--Whitney~U tests applied to the full AIC and
BIC distributions (Table~\ref{tab:aic_bic_tests} in
~\ref{app:tables}) provide a complementary perspective. For \texttt{1d},
{both tests are formally significant ($D = 0.098$,
$p_{\rm KS} = 2.42\times10^{-5}$, $p_{\rm MW} = 4.96\times10^{-2}$),
so the joint criterion is satisfied, indicating a statistically
significant difference between repeaters and non-repeaters.} For
\texttt{cavity}, {a similar result is obtained ($D = 0.101$,
$p_{\rm KS} = 1.10\times10^{-5}$, $p_{\rm MW} = 3.61\times10^{-2}$).}
For the \texttt{point} model, {both tests also yield significant results
($D = 0.100$, $p_{\rm KS} = 1.44\times10^{-5}$,
$p_{\rm MW} = 3.59\times10^{-2}$), indicating evidence for a difference
between repeaters and non-repeaters. However, the associated effect
sizes remain small ($D\simeq0.10$), indicating limited practical
separation between the two classes.}

{For the extended five-model comparison, the \texttt{1d}
model gives $D=0.098$, $p_{\rm KS}=2.42\times10^{-5}$, and
$p_{\rm MW}=4.96\times10^{-2}$; \texttt{cavity} gives $D=0.101$,
$p_{\rm KS}=1.10\times10^{-5}$, and $p_{\rm MW}=3.61\times10^{-2}$;
\texttt{point} gives $D=0.100$, $p_{\rm KS}=1.44\times10^{-5}$, and
$p_{\rm MW}=3.59\times10^{-2}$; \texttt{sinc2} gives $D=0.098$,
$p_{\rm KS}=4.85\times10^{-5}$, and $p_{\rm MW}=6.26\times10^{-2}$;
and \texttt{powerlaw} gives $D=0.097$, $p_{\rm KS}=5.27\times10^{-5}$,
and $p_{\rm MW}=6.59\times10^{-2}$. Of the five models, \texttt{point},
\texttt{1d}, and \texttt{cavity} satisfy the joint significance criterion,
whereas \texttt{sinc2} and \texttt{powerlaw} do not. The corresponding
effect sizes remain small across all five models ($D\simeq0.097$--$0.101$),
indicating limited practical separation between repeaters and
non-repeaters despite the formal statistical significance obtained for
three of the five templates. These results are computed after both the
quality cut and the common-sample restriction, ensuring that the
repeater/non-repeater comparisons are performed on the same set of FRBs
across all models.}

{For the BIC distributions, the statistical results are
numerically essentially identical to those obtained for AIC:
\texttt{point}, \texttt{1d}, and \texttt{cavity} satisfy the joint
criterion, whereas \texttt{sinc2} and \texttt{powerlaw} do not.}

{Overall, the extended five-model comparison, restricted
to the common quality-selected sample, shows that the \texttt{1d}
curvature-radiation template remains a statistically competitive
description of the observed spectra, but its preference within the
original curvature-radiation-inspired family does not imply a unique
preference for curvature radiation itself. The non-curvature
$\mathrm{sinc}^{2}$ baseline is the single most frequently preferred
model under both AIC and BIC, for both repeaters and non-repeaters.
These results indicate that a substantial part of the spectral structure
captured by the \texttt{1d} model is likely associated with the
oscillatory modulation rather than uniquely with the curvature-radiation
spectral envelope. Consequently, the AIC/BIC analysis is interpreted here
as a comparison of the relative adequacy of the tested spectral
templates, rather than as independent evidence establishing curvature
radiation as the underlying emission or propagation mechanism.}

\subsection{Ljung--Box Residual Autocorrelation Test}

\begin{figure*}[ht]
    \centering
    \includegraphics[width=\textwidth]{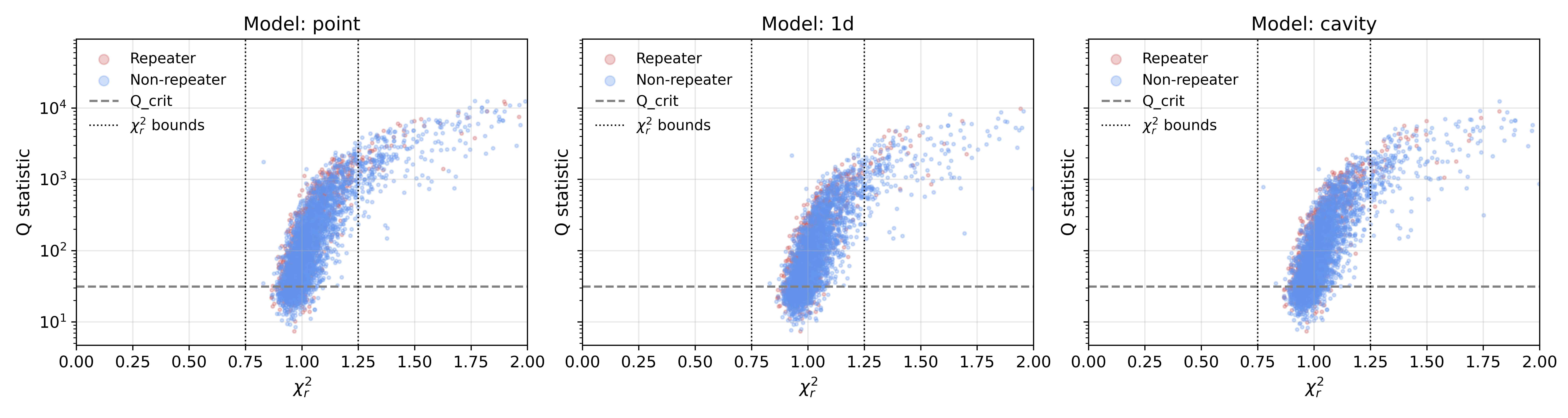}
    \caption{Scatter plot of the Ljung--Box statistic $Q$ versus reduced chi-squared $\chi^2_r$ for the \texttt{point} (a), \texttt{1d} (b), and \texttt{cavity} (c) models, coloured by FRB class (repeaters in red, non-repeaters in blue). The dashed vertical lines mark the boundaries of the region $0.75 \leq \chi^2_r \leq 1.25$ and 
    the horizontal dashed line marks the LB pass threshold. Sources passing both criteria simultaneously occupy the lower-left region of each panel.}
    \label{fig:lb_scatter}
\end{figure*}

The Ljung--Box (LB) Q test evaluates the null hypothesis that residuals are independently distributed, i.e., exhibit no significant autocorrelation up to a specified lag. A model is considered to pass the test when $p_{\rm LB}$ exceeds a chosen significance level (typically $p_{\rm LB} > 0.05$) \citep{Mokeddem2023}. Figure~\ref{fig:lb_scatter} shows the scatter of the LB statistic $Q$ versus $\chi^2_r$ for each model, coloured by FRB class. Table~\ref{tab:lb_passrates} summarises the pass rates and Table~\ref{tab:lb_statistics} the full descriptive statistics for $Q$, $Q/{\rm dof}$, and $p_{\rm LB}$.

Pass rates for the LB test alone are considerably lower than those for $\chi^2_r$, ranging from $14.6\%$ to $21.6\%$ for non-repeaters and $12.1\%$ to $19.5\%$ for repeaters. The \texttt{1d} model achieves the highest LB pass rates in both classes ($21.6\%$ and $19.5\%$), followed by \texttt{cavity} ($19.7\%$ and $15.7\%$) and \texttt{point} ($14.6\%$ and $12.1\%$). When both $\chi^2_r$ and LB criteria are required simultaneously, pass rates remain identical to the LB pass rates: \texttt{1d} retains $21.6\%$ of non-repeaters and $19.5\%$ of repeaters, while \texttt{point} yields the lowest joint pass rates ($14.6\%$ and $12.1\%$, respectively). The exact equality between the Ljung–Box and joint pass rates indicates that all fits satisfying the residual-independence criterion also satisfy the $\chi^2_r$ quality threshold, suggesting that residual autocorrelation constitutes an important limitation of the current templates.

\begin{table*}[ht]
    \centering
    \caption{Pass rates for $\chi^2_r$, Ljung--Box, and both criteria 
    simultaneously, by model and FRB class.}
    \label{tab:lb_passrates}
    \begin{tabular}{llccccc}
        \hline
        Model & Class & $N$ & Pass $\chi^2_r$ & Pass LB & Pass both \\
        \hline
        \texttt{point}  
            & Non-rep. & 3555 & 3208 (90.2\%) & 520 (14.6\%) & 520 (14.6\%) \\
            & Rep.     &  981 &  925 (94.3\%) & 119 (12.1\%) & 119 (12.1\%) \\
        \texttt{1d}     
            & Non-rep. & 3554 & 3295 (92.7\%) & 769 (21.6\%) & 769 (21.6\%) \\
            & Rep.     &  981 &  942 (96.0\%) & 191 (19.5\%) & 191 (19.5\%) \\
        \texttt{cavity} 
            & Non-rep. & 3555 & 3275 (92.1\%) & 699 (19.7\%) & 699 (19.7\%) \\
            & Rep.     &  981 &  937 (95.5\%) & 154 (15.7\%) & 154 (15.7\%) \\
        \hline
    \end{tabular}
\end{table*}

The median $p$-values are extremely low across all models and classes ($\tilde{p}_{\rm LB} \lesssim 10^{-7}$), indicating that residual autocorrelation is pervasive. Repeaters consistently show lower median $p$-values than non-repeaters in all models, indicating stronger residual autocorrelation in repeating bursts.  Interestingly, although repeaters exhibit narrower $\chi^2_r$  distributions, this result shows that good global fit quality does not necessarily imply more independent residual structure. The normalized statistic $\tilde{Q}/{\rm dof}$ (Table~\ref{tab:lb_statistics}) follows the same ordering across models: \texttt{point} yields the largest values ($5.43$ for non-repeaters and $6.53$ for repeaters), while \texttt{1d} yields the smallest ($3.50$ and $3.90$, respectively), consistent with its superior LB pass rates. The heavy tails in both $Q$ and $Q/{\rm dof}$ (max $\sim 10^4$ and $\sim 10^2$, respectively) indicate a subset of fits with strongly correlated residuals that pull the means well above the medians. In all cases, the mean values exceed the medians by nearly an order of magnitude, reinforcing the strongly right-skewed nature of the residual-autocorrelation distributions.

Taken together, the LB results indicate that residual autocorrelation remains a significant limitation of the current templates. Joint pass rates of $12.1$--$21.6\%$ indicate that most spectra retain statistically significant residual correlations after model subtraction. This does not invalidate the $\chi^2_r$ goodness-of-fit results, since $\chi^2_r$ and the LB test probe different aspects of the residuals. Instead, it indicates that the tested templates capture the dominant spectral envelope but do not exhaust the spectral complexity of most CHIME/FRB spectra. Among the three models, the \texttt{1d} template provides the most favourable residual-independence performance.

\subsection{Curvature-Radiation Model Parameters}
\label{sec:model_parameters}

Table~\ref{tab:param_summary} summarizes the fitted parameters $s_0$ and $b$ 
for each model after applying the quality cut, and Table~\ref{tab:param_tests} 
presents the KS and Mann--Whitney U tests comparing repeaters and 
non-repeaters.

For the \texttt{point} model, the parameter $s_0$ remains extremely small for both populations, with median values $\tilde{s}_0 \approx 5\times10^{-6}$ and narrow distributions concentrated near zero. The KS and MW tests both indicate no statistically significant difference between repeaters and non-repeaters ($p_{\rm KS} \approx 1.42\times10^{-1}$ and $p_{\rm MW} \approx 1.69 \times 10^{-1}$). Given the extremely small absolute parameter values and the absence of an interference term, small numerical differences in $s_0$ are not interpreted as physically meaningful. The parameter $b$ is undefined in the \texttt{point} model and is therefore excluded from the statistical analysis.

For the \texttt{1d} model, both $s_0$ and $b$ exhibit highly significant differences between repeaters and non-repeaters according to both the KS and Mann--Whitney tests ($p \ll 0.05$). This model also yields the largest KS statistics among the three models ($D = 0.090$ for $s_0$ and $D = 0.084$ for $b$), indicating the strongest population-level separation observed in the parameter space. Repeaters show slightly larger median values of both parameters, with $\tilde{s}_0 = 1.96\times10^{-4}$ compared to $1.63\times10^{-4}$ for non-repeaters and $\tilde{b} = 1.89\times10^{-3}$ compared to $1.74\times10^{-3}$, respectively. However, the broad overlap between the interquartile ranges indicates that these differences remain modest despite their statistical significance. 

The \texttt{cavity} model presents a more nuanced behaviour. The interference 
scale $b$ shows a mild but statistically non-significant offset between repeaters and non-repeaters 
($D = 0.044$, $p_{KS} \approx 1.09\times10^{-1}$), whereas the $s_0$ distributions 
are also statistically indistinguishable according to both the KS and 
Mann--Whitney tests ($p_{\rm KS} \approx 6.80\times10^{-1}$ and 
$p_{\rm MW} \approx 8.37 \times 10^{-1}$, respectively). In contrast to the \texttt{1d} case, the cavity model shows no statistically robust population-level separation in either $s_0$ or $b$. The strongly overlapping quartile ranges and large variances indicate that dispersion dominates over any systematic shift between the two populations. 

 \begin{table*}[ht]
    \centering
    \caption{Descriptive statistics for the fitted parameters $s_0$ and $b$ 
    by model and FRB class (quality cut applied). 
    $\bar{x}$ and $\tilde{x}$ denote mean and median, respectively, while 
    $x_{25}$ and $x_{75}$ represent the 25th and 75th percentiles.}
    \label{tab:param_summary}
    \begin{tabular}{lllllrrrr}
        \hline
        Model & Param & Class & $N$ & $\bar{x}$ & $\tilde{x}$ & $\sigma$ 
              & $x_{25}$ & $x_{75}$ \\
        \hline

        \texttt{point} 
            & $s_0$ & Non-rep. & 3\,208 
            & $7.36\times10^{-6}$ 
            & $4.77\times10^{-6}$ 
            & $9.16\times10^{-6}$ 
            & $1.14\times10^{-6}$ 
            & $9.87\times10^{-6}$ \\

            &       & Rep. & 925 
            & $7.11\times10^{-6}$ 
            & $5.27\times10^{-6}$ 
            & $7.41\times10^{-6}$ 
            & $1.43\times10^{-6}$ 
            & $1.05\times10^{-5}$ \\

        \hline

        \texttt{1d}
            & $s_0$ & Non-rep. & 3\,295 
            & $2.84\times10^{-4}$ 
            & $1.63\times10^{-4}$ 
            & $7.51\times10^{-4}$ 
            & $7.62\times10^{-5}$ 
            & $3.22\times10^{-4}$ \\

            &       & Rep. & 942 
            & $3.28\times10^{-4}$ 
            & $1.96\times10^{-4}$ 
            & $6.94\times10^{-4}$ 
            & $9.09\times10^{-5}$ 
            & $4.06\times10^{-4}$ \\

            & $b$ & Non-rep. & 3\,295 
            & $2.86\times10^{-3}$ 
            & $1.74\times10^{-3}$ 
            & $7.90\times10^{-3}$ 
            & $1.42\times10^{-3}$ 
            & $2.35\times10^{-3}$ \\

            &     & Rep. & 942 
            & $2.37\times10^{-3}$ 
            & $1.89\times10^{-3}$ 
            & $4.89\times10^{-3}$ 
            & $1.46\times10^{-3}$ 
            & $2.44\times10^{-3}$ \\

        \hline

        \texttt{cavity}
            & $s_0$ & Non-rep. & 3\,275 
            & $2.46\times10^{-4}$ 
            & $1.57\times10^{-5}$ 
            & $8.07\times10^{-4}$ 
            & $1.92\times10^{-6}$ 
            & $4.23\times10^{-5}$ \\

            &       & Rep. & 937 
            & $2.69\times10^{-4}$ 
            & $1.71\times10^{-5}$ 
            & $8.41\times10^{-4}$ 
            & $8.18\times10^{-14}$ 
            & $3.98\times10^{-5}$ \\

            & $b$ & Non-rep. & 3\,275 
            & $4.56\times10^{-3}$ 
            & $3.00\times10^{-3}$ 
            & $1.48\times10^{-2}$ 
            & $1.00\times10^{-4}$ 
            & $4.03\times10^{-3}$ \\

            &     & Rep. & 937 
            & $4.29\times10^{-3}$ 
            & $2.72\times10^{-3}$ 
            & $1.42\times10^{-2}$ 
            & $1.00\times10^{-4}$ 
            & $4.03\times10^{-3}$ \\

        \hline
    \end{tabular}
\end{table*}

In both the \texttt{1d} and \texttt{cavity} models, the distributions are strongly right-skewed, making the median and interquartile range more representative than the mean and standard deviation. The fitted parameters reinforce the conclusion that the \texttt{1d} model provides the most pronounced statistical differences between repeaters and non-repeaters among the models considered. Nevertheless, the substantial overlap between the parameter distributions suggests that these differences should be interpreted as gradual population trends rather than evidence for a distinct physical dichotomy. 

\subsection{{Effective Modulation Scales} from the Fitted Parameters}
\label{sec:physical_constraints}

As discussed in Section \ref{sec:fitting}, the {effective modulation timescale} $b$ can be mapped onto {an effective spatial scale} through {$L_{\rm eff} = cb/\pi$}. The median inferred scales (Table~\ref{tab:physical_constraints}) are $\tilde{L}\approx16.5$--$17.9$~cm for the \texttt{1d} model and $\tilde{L}\approx25$--$28$~cm for the \texttt{cavity} model.  These values fall within the same order of magnitude as the compact magnetospheric scales typically considered in coherent curvature-radiation models ~\citep{Yang2020,Cui2023}, although they are systematically larger than the nominal theoretical estimates by factors of roughly $1.5$--$2.5$.

The inferred scales show substantial overlap between repeaters and non-repeaters. In the \texttt{1d} model, repeaters favour slightly larger characteristic lengths than non-repeaters ($\Delta L \approx 1.4$~cm), while the opposite trend appears in the \texttt{cavity} model ($\Delta L \approx 3$~cm). However, these offsets are small compared with the  widths of the distributions, indicating that both populations occupy largely overlapping regions of parameter space.

{The identification of $L_{\rm eff}$ with a bunch length or a cavity separation is conditional on the assumed emission model. Similar spectral modulation may also arise from propagation effects, including scintillation and plasma lensing, or from unresolved frequency-dependent temporal structure within the burst. The band-averaged spectral fits performed here do not allow these possibilities to be distinguished from intrinsic spectral modulation. We therefore interpret $b$ conservatively as an effective modulation parameter rather than as a direct measurement of a physical bunch length or cavity separation. We further note that the present data do not measure the particle Lorentz factor, the field-line curvature radius, or the emission altitude: as discussed in Section \ref{sec:fitting}, $\nu_c$ is not independently identifiable within $400$--$800$ MHz and is nearly degenerate with the amplitude normalisation. The modulation parameter $b$ is one of the few geometrical combinations that remain accessible from the normalized, band-limited spectra used in this study, which is the reason it is reported here.}

Overall, the fitted {effective modulation scales} are broadly consistent with the compact emission regions expected in coherent curvature-radiation scenarios{, under the conditional interpretation adopted above}, but they do not support a clear physical separation between repeating and non-repeating FRBs. Determining whether the observed population-level trends reflect intrinsic source properties, modelling assumptions, or observational selection effects will require broader frequency coverage and more detailed physical modelling.

\section{Conclusion}\label{sec:conclusions}

We have presented a population-level spectral analysis of 4\,536 FRBs from the second CHIME/FRB catalog using three curvature-radiation models representing progressively richer interference geometries — point-source, one-dimensional bunch, and paired-bunch cavity. Although all models yield reduced chi-squared values clustered near unity across a broad range of spectral binning factors, residual autocorrelation
diagnosed by the Ljung--Box test emerges as an important limitation: only $12\%$--$22\%$ of bursts simultaneously satisfy both the reduced-$\chi^{2}$ and Ljung--Box criteria, indicating that the tested templates capture the dominant spectral envelope but do not exhaust the spectral complexity of most CHIME/FRB spectra. This behaviour does not argue against a curvature-radiation origin; rather, it suggests that additional physical ingredients, such as more complex bunch geometries, intrinsic spectral evolution, or propagation-induced modulation, contribute to the observed spectral structure. Repeaters consistently exhibit slightly better fit quality than non-repeaters, although the differences remain modest (KS statistics $D\approx0.080$--$0.103$) and do not imply a sharp physical dichotomy. Thus, the main limitation of the present models lies not in reproducing the broad spectral envelope, but in describing the full fine-scale spectral structure present in many bursts. 

Among the models considered, the \texttt{1d} geometry provides the most successful overall description of the data. It achieves the highest pass rates under both the reduced-$\chi^{2}$ and Ljung--Box diagnostics and is the model most frequently preferred by AIC and BIC on a burst-by-burst basis ($61\%$--$70\%$ under AIC; $46\%$--$57\%$ under BIC). However, the information criteria provide only weak discrimination between repeaters and non-repeaters, indicating that both populations share broadly similar spectral phenomenology within the framework explored here. {Since this comparison is restricted to templates within the same curvature-radiation-inspired family, it identifies the preferred template within that family and does not by itself favour curvature radiation over competing emission or propagation mechanisms. {When the two non-curvature baseline models are included, this ranking changes.
The {\texttt{sinc2}} model becomes the single most frequently preferred template under
both AIC ({$38.8\%$} of non-repeaters, {$38.7\%$} of repeaters) and BIC ({$45.1\%$} of
non-repeaters, {$44.6\%$} of repeaters), with the 1d model second under both
criteria ({$29.1\%$}/{$32.1\%$} under AIC; {$21.9\%$}/{$22.6\%$} under BIC). Once
phenomenological oscillatory baselines are considered, the data therefore show
no preference for the curvature-radiation spectral envelope specifically: the
{oscillatory spectral structure is more strongly supported than the curvature-radiation envelope}.}}

The fitted {effective modulation scales} correspond to characteristic {effective spatial scales} of approximately $\tilde{L}\approx16.5$--$17.9$\,cm in the \texttt{1d} model and $\tilde{L}\approx25$--$28$\,cm in the \texttt{cavity} model. These scales are broadly consistent with the compact emission regions expected in coherent curvature-radiation scenarios, although they exceed nominal theoretical estimates by factors of roughly $1.5$--$2.5$~\citep{Yang2020,Cui2023}. {Their identification with bunch lengths or cavity separations remains conditional on the assumed emission model, since comparable spectral modulation may also arise from propagation effects or from unresolved temporal structure (Section \ref{sec:physical_constraints}).} Statistically significant differences between repeaters and non-repeaters are detected in the \texttt{1d} model, but the strongly overlapping parameter distributions ($D\le0.090$) indicate that these trends remain modest and do not constitute evidence for a distinct physical dichotomy.

{Overall, our results show that the tested curvature-radiation-inspired templates provide useful phenomenological descriptions of a subset of FRB spectra, but they do not uniquely identify coherent curvature radiation as the underlying emission mechanism. They also demonstrate} the limitations of {the} simplest implementations {of this framework}, particularly in capturing persistent fine-scale spectral structure revealed by residual diagnostics. The spectral properties of repeaters and non-repeaters are broadly consistent with a common underlying physical origin, although modest population-level differences remain detectable within the  \texttt{1d} model. Future progress will require broadband observations capable of constraining the spectral cutoff frequency directly, together with physically richer curvature-radiation models that connect the phenomenological {modulation} scale to specific plasma processes. Repeated-burst analyses and realistic simulations will be particularly important for determining whether the modest population-level trends identified here arise from intrinsic source physics or from observational and processing effects.

\section*{Acknowledgements}
JASF acknowledges support from the National Research Foundation
of South Africa. WSHR acknowledges partial support from FAPES (Brazil) and CNPq (Brazil). TVA acknowledges Prof. Leonardo D. Secchin for kindly providing access to the computational resources used in the calculations presented in this work. These resources were acquired with support from FAPES (grant nº 116/2019).

\section*{Data availability}

The data underlying this study are publicly available as part of the second CHIME/FRB public data release at the CHIME/FRB catalog: \url{https://www.chime-frb.ca/catalog2}.

\section*{Declaration of competing interest}

The authors declare that they have no known competing financial interests or personal relationships that could have appeared to influence the work reported in this paper.

\bibliographystyle{elsarticle-harv}
\bibliography{frb-curvature}

\newpage

\appendix
\section{Single-Particle Curvature Spectrum}\label{appendix:single_particle}

The energy radiated per unit frequency interval per unit solid angle by a relativistic charged particle in instantaneous circular motion is given by~\citep{Jackson1999}:
\begin{equation}
    \frac{d^2I}{d\nu  \, d\Omega} = \frac{8 \pi e^2 \nu^2 \rho^2}{3 c^3}
    \left(\frac{1}{\gamma^2} + \theta^2\right)^2 
    \left[ K_{2/3}^2(\xi_\nu) + \frac{\theta^2}{\theta^2+(1/\gamma^2)} K_{1/3}^2(\xi_\nu) \right],
\end{equation}
where the argument of the modified Bessel functions $K_\alpha$ is
\begin{equation}
    \xi_\nu = \frac{2\pi \nu \rho}{3c} \left(\frac{1}{\gamma^2} + \theta^2\right)^{3/2}.
\end{equation}

For $\theta = 0$ and $\xi_\nu = 1/2$, the critical frequency $\nu_c = 3c\gamma^3/(4\pi\rho)$ defines the spectral turnover. Using the asymptotic expansions of $K_\alpha(\xi_\nu)$, the radiation behavior is:
\begin{align}
    \frac{d^2I}{d\nu  \, d\Omega}\bigg|_{\nu \ll \nu_c} 
    &\approx \frac{3}{2^{1/3}}\frac{[e\gamma \Gamma(2/3)]^2}{\pi c}
    \left(\frac{\nu}{\nu_c}\right)^{2/3}, \\[6pt]
    \frac{d^2I}{d\nu  \, d\Omega}\bigg|_{\nu \gg \nu_c} 
    &\approx \frac{3 e^2 \gamma^2}{2c}
    \frac{\nu}{\nu_c} e^{-\nu/\nu_c},
\end{align}
where $\Gamma$ denotes the Euler gamma function. Combining both asymptotic forms yields a practical analytic representation for the full spectrum~\citep{Yang2018}:
\begin{equation}
    \frac{d^2I}{d\nu  \, d\Omega} \approx 
    A\,\gamma^2 \left(\frac{\nu}{\nu_c}\right)^{2/3} e^{-\nu/\nu_c},
\end{equation}
with $A = (3/2^{1/3})[e\,\Gamma(2/3)]^2/(\pi c)$. This formulation forms the foundation of the multi-charge models discussed in Section~\ref{sec:radiation}.

\section{Supplementary Statistical Tables and Diagnostics}
\label{app:tables}

This appendix reports supplementary statistical tests and diagnostic checks supporting the population-level comparisons discussed in Section~\ref{sec:results}. These include two-sample KS and Mann--Whitney U tests, detailed Ljung--Box statistics, fitted-parameter comparisons, and additional checks related to masking patterns in the residual autocorrelation analysis.

\begin{table}[ht]
    \centering
    \caption{KS and Mann--Whitney U tests (repeaters vs.\ non-repeaters)  for $\chi^2_r$ distribution at $f_{\rm bin} = 4$.}
    \label{tab:stat_tests}
    \begin{tabular}{lcccc}
        \hline
        Model & $D$ & $p_{\rm KS}$ & $U$ & $p_{\rm MW}$ \\
        \hline
        \texttt{1d}     & 0.103 & $1.26\times10^{-7}$  & 1\,521\,151 & $9.50\times10^{-10}$ \\
        \texttt{cavity} & 0.083 & $4.64\times10^{-5}$  & 1\,572\,984 & $2.57\times10^{-6}$  \\
        \texttt{point}  & 0.080 & $1.11\times10^{-4}$  & 1\,577\,239 & $4.54\times10^{-6}$  \\
        \hline
    \end{tabular}
\end{table}

\begin{table*}[ht]
    \centering
    \caption{KS and Mann--Whitney~U tests (repeaters vs.\ non-repeaters) for AIC and BIC distributions with quality cut applied. The column ``Significant'' flags cases where both $p_{\rm KS}$ and $p_{\rm MW}$ satisfy $p<0.05$.}
    \label{tab:aic_bic_tests}
    \begin{tabular}{llccccl}
        \hline
        Criterion & Model & $D$ & $p_{\rm KS}$
                  & $U$ & $p_{\rm MW}$ & Significant \\
        \hline
        AIC & \texttt{point}
                & $0.100$ & $1.44\times10^{-5}$
                & $1\,027\,631$ & $3.59\times10^{-2}$ & YES \\
            & \texttt{1d}
                & $0.098$ & $2.42\times10^{-5}$
                & $1\,024\,505$ & $4.96\times10^{-2}$ & YES \\
            & \texttt{cavity}
                & $0.101$ & $1.10\times10^{-5}$
                & $1\,027\,581$ & $3.61\times10^{-2}$ & YES \\
            & \texttt{sinc2}
                & $0.098$ & $4.85\times10^{-5}$
                & $960\,789$ & $6.26\times10^{-2}$ & NO \\
            & \texttt{powerlaw}
                & $0.097$ & $5.27\times10^{-5}$
                & $960\,281$ & $6.59\times10^{-2}$ & NO \\
        BIC & \texttt{point}
                & $0.100$ & $1.44\times10^{-5}$
                & $1\,027\,631$ & $3.59\times10^{-2}$ & YES \\
            & \texttt{1d}
                & $0.098$ & $2.42\times10^{-5}$
                & $1\,024\,497$ & $4.96\times10^{-2}$ & YES \\
            & \texttt{cavity}
                & $0.101$ & $1.10\times10^{-5}$
                & $1\,027\,584$ & $3.61\times10^{-2}$ & YES \\
            & \texttt{sinc2}
                & $0.098$ & $4.85\times10^{-5}$
                & $960\,784$ & $6.27\times10^{-2}$ & NO \\
            & \texttt{powerlaw}
                & $0.097$ & $5.27\times10^{-5}$
                & $960\,283$ & $6.59\times10^{-2}$ & NO \\
        \hline
    \end{tabular}
\end{table*}

\begin{table*}[ht]
    \centering
    \caption{Descriptive statistics for the Ljung--Box statistic $Q$, 
    normalised statistic $Q/{\rm dof}$, and $p$-value, by model and FRB class. 
    $\bar{x}$ denotes the mean, $\tilde{x}$ the median, while 
    $x_{25}$ and $x_{75}$ represent the 25th and 75th percentiles.}
    \label{tab:lb_statistics}
    \begin{tabular}{lllrrrrrrr}
        \hline
        Statistic & Model & Class & $N$ & $\bar{x}$ & $\sigma$ & $\tilde{x}$ & $x_{25}$ & $x_{75}$ \\
        \hline
        $Q$ 
            & \texttt{1d}     
                & Non-rep. & 3554 & 567.3 & 2822.9 &  70.0 &  33.8 & 211.2 \\
            &                 
                & Rep.     &  981 & 363.5 & 1363.8 &  78.1 &  37.4 & 242.5 \\
            & \texttt{cavity} 
                & Non-rep. & 3555 & 630.1 & 3055.3 &  80.2 &  35.8 & 261.0 \\
            &                 
                & Rep.     &  981 & 480.6 & 1780.2 &  99.5 &  42.9 & 312.0 \\
            & \texttt{point}  
                & Non-rep. & 3555 & 852.0 & 3392.2 & 108.6 &  43.6 & 360.3 \\
            &                 
                & Rep.     &  981 & 705.2 & 2544.3 & 130.6 &  52.4 & 434.4 \\
        \hline
        $Q/{\rm dof}$ 
            & \texttt{1d}     
                & Non-rep. & 3554 & 28.36 & 141.15 &  3.50 &  1.69 & 10.56 \\
            &                 
                & Rep.     &  981 & 18.17 &  68.19 &  3.90 &  1.87 & 12.13 \\
            & \texttt{cavity} 
                & Non-rep. & 3555 & 31.50 & 152.77 &  4.01 &  1.79 & 13.05 \\
            &                 
                & Rep.     &  981 & 24.03 &  89.01 &  4.97 &  2.15 & 15.60 \\
            & \texttt{point}  
                & Non-rep. & 3555 & 42.60 & 169.61 &  5.43 &  2.18 & 18.01 \\
            &                 
                & Rep.     &  981 & 35.26 & 127.22 &  6.53 &  2.62 & 21.72 \\
        \hline
         $p_{\rm LB}$ 
            & \texttt{1d}     
                & Non-rep. & 3554 & $7.11\times10^{-2}$ & $1.71\times10^{-1}$ & $1.82\times10^{-7}$ & $6.92\times10^{-34}$ & $2.77\times10^{-2}$ \\
            &                 
                & Rep.     &  981 & $6.16\times10^{-2}$ & $1.60\times10^{-1}$ & $8.25\times10^{-9}$ & $3.65\times10^{-40}$ & $1.04\times10^{-2}$ \\
            & \texttt{cavity} 
                & Non-rep. & 3555 & $6.30\times10^{-2}$ & $1.61\times10^{-1}$ & $3.67\times10^{-9}$ & $6.91\times10^{-44}$ & $1.64\times10^{-2}$ \\
            &                 
                & Rep.     &  981 & $4.70\times10^{-2}$ & $1.38\times10^{-1}$ & $1.56\times10^{-12}$ & $2.84\times10^{-54}$ & $2.11\times10^{-3}$ \\
            & \texttt{point}  
                & Non-rep. & 3555 & $4.41\times10^{-2}$ & $1.33\times10^{-1}$ & $3.54\times10^{-14}$ & $3.44\times10^{-64}$ & $1.69\times10^{-3}$ \\
            &                 
                & Rep.     &  981 & $3.63\times10^{-2}$ & $1.24\times10^{-1}$ & $3.05\times10^{-18}$ & $1.44\times10^{-79}$ & $9.98\times10^{-5}$ \\
        \hline
    \end{tabular}
\end{table*}

\begin{table*}[ht]
    \centering
    \caption{KS and Mann--Whitney~U tests comparing repeater and non-repeater 
    distributions of fitted parameters $s_0$ and $b$ (quality cut applied). 
    The column ``Significant'' flags cases where both 
    $p_{\rm KS}$ and $p_{\rm MW}$ satisfy $p < 0.05$.}
    \label{tab:param_tests}
    \begin{tabular}{llccccl}
        \hline
        Model & Param & $D$ & $p_{\rm KS}$ & $U$ & $p_{\rm MW}$ & Significant \\
        \hline
        \texttt{point}  & $s_0$
            & $0.043$ & $1.42\times10^{-1}$
            & $1\,527\,668$ & $1.69\times10^{-1}$ & NO \\
        \texttt{1d}     & $s_0$
            & $0.090$ & $1.44\times10^{-5}$
            & $1\,712\,302$ & $1.28\times10^{-6}$ & YES \\
                        & $b$
            & $0.084$ & $5.45\times10^{-5}$
            & $1\,666\,059$ & $5.68\times10^{-4}$ & YES \\
        \texttt{cavity} & $s_0$
            & $0.026$ & $6.80\times10^{-1}$
            & $1\,541\,070$ & $8.37\times10^{-1}$ & NO \\
                        & $b$
            & $0.044$ & $1.09\times10^{-1}$
            & $1\,495\,066$ & $2.31\times10^{-1}$ & NO \\
        \hline
    \end{tabular}
\end{table*}

\begin{table*}[ht]
    \centering
    \caption{Statistics of the {effective} spatial scale {$L_{\rm eff} = cb/\pi$} inferred from the fitted values of $b$ for the \texttt{1d} and \texttt{cavity} models (quality cut applied). $L$ is given in cm, using $c = 3\times10^{10}$cm\~s$^{-1}$. The quantities $L_{25}$ and $L_{75}$ denote the 25th and 75th percentiles, respectively. {The interpretations listed in the last column are conditional on the assumed emission model (see Section \ref{sec:physical_constraints}).}}
    \label{tab:physical_constraints}
    \begin{tabular}{llcccccl}
        \hline
        Model & Class & $\bar{L}$ & $\tilde{L}$ & $\sigma_L$
              & $L_{25}$ & $L_{75}$ & Interpretation \\
              &       & (cm) & (cm) & (cm) & (cm) & (cm) & \\
        \hline
        \texttt{1d}
            & Non-rep.
            & $18.05$
            & $16.46$
            & $6.92$
            & $13.56$
            & $22.04$
            & Bunch length $l$ {(conditional)} \\
            & Rep.
            & $18.89$
            & $17.90$
            & $7.24$
            & $13.94$
            & $23.09$
            & Bunch length $l$ {(conditional)} \\
        \texttt{cavity}
            & Non-rep.
            & $21.94$
            & $27.82$
            & $17.56$
            & $0.95$
            & $37.67$
            & Cavity separation $d$ {(conditional)} \\
            & Rep.
            & $21.04$
            & $24.98$
            & $17.55$
            & $0.95$
            & $38.00$
            & Cavity separation $d$ {(conditional)} \\
        \hline
    \end{tabular}
\end{table*}

\newpage

\begin{figure*}[htb!]
    \centering
    \includegraphics[width=0.9\textwidth]{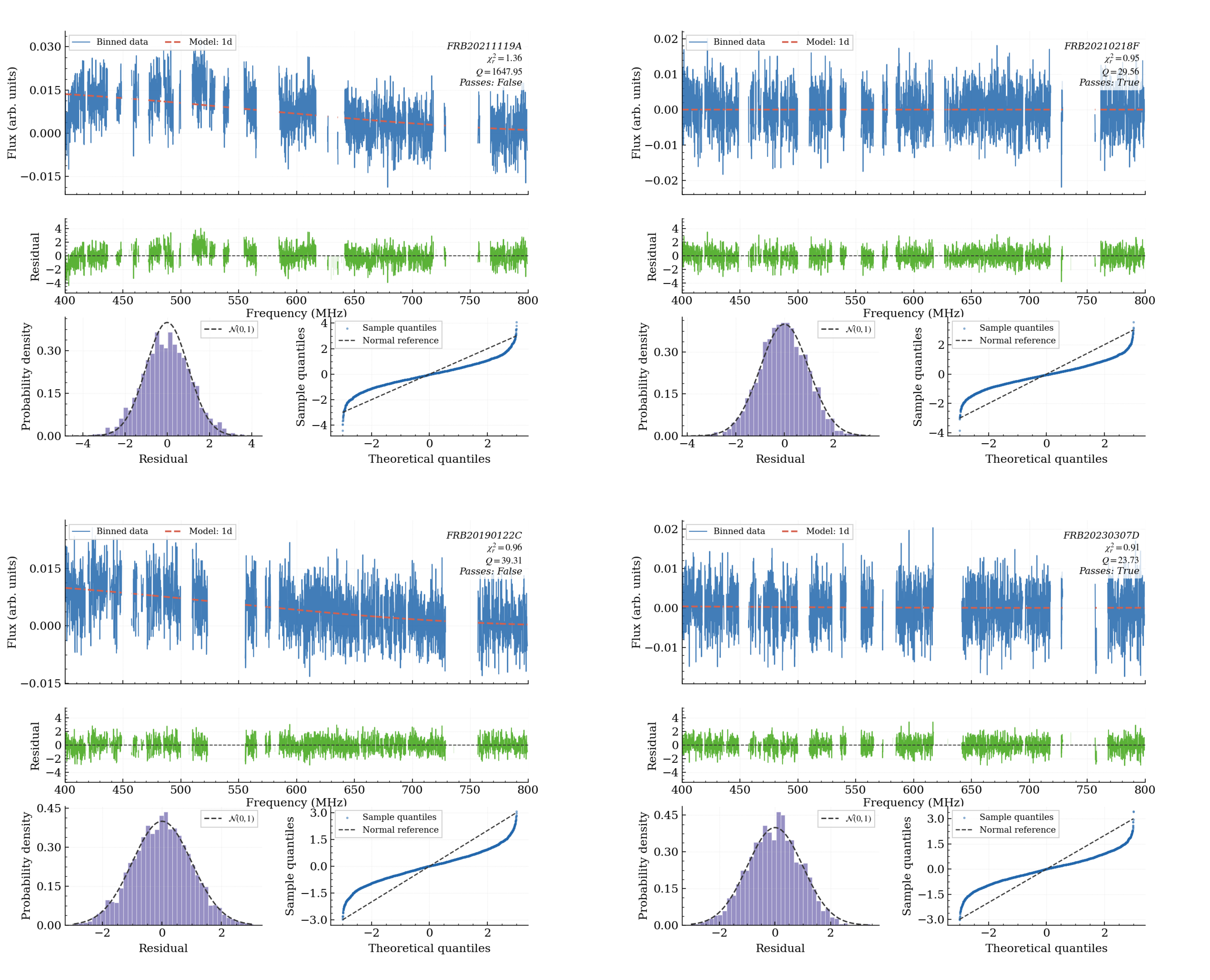}
    \caption{Spectral fit diagnostics for four representative FRBs detected by CHIME, fitted with a curvature radiation \texttt{1d} model over the frequency range $400$--$800$\,MHz.
    Each sub-figure contains four panels.\textit{Top:} Binned spectrum (blue), for $f_{\rm bin} = 4$, and best-fit model (dashed red); \textit{Upper middle:} Weighted residuals.\textit{Lower left:} Probability density of the weighted residuals (purple) overlaid with a standard normal distribution $\mathcal{N}(0,1)$ (dashed black), serving as a visual goodness-of-fit diagnostic;
    \textit{Lower right:} Quantile--quantile (Q--Q) plot comparing the empirical distribution of residuals against Gaussian theoretical quantiles; deviation from the dashed reference line indicates departures from normality. Flux densities are given in arbitrary units as described on Section~\ref{sec:preprocessing}.}
    \label{fig:spectral_fits}
\end{figure*}

\end{document}